\documentclass[aps,pre,twocolumn,floatfix,nofootinbib,showpacs,groupedaddress,superscriptaddress]{revtex4}
\usepackage[sort&compress]{natbib}
\usepackage{graphicx}
\usepackage{epsfig}

\begin{document}
\title{Long DNA molecule as a pseudoscalar liquid crystal}
\author{K.G. Petrosyan}
\affiliation{Institute of Physics, Academia Sinica, Nankang, Taipei
11529, Taiwan}
\author{Chin-Kun Hu}
\affiliation{Institute of Physics, Academia Sinica, Nankang,
Taipei 11529, Taiwan} \affiliation{Center for Nonlinear and
Complex Systems and Department of Physics, Chung-Yuan Christian
University, Chungli, 320 Taiwan}
\date{\today}

\begin{abstract}
We show that a long DNA molecule can form a novel condensed phase
of matter, the {\it pseudoscalar liquid crystal}, that consists of
aperiodically ordered DNA fragments in right-handed $B$ and left-handed $Z$ forms. We
discuss the possibility of transformation of $B$-DNA into $Z$-DNA
and vice versa via first-order phase transitions as well as
transformations from the phase with zero total chirality into pure
$B$- or $Z$-DNA samples through second-order phase transitions.
The presented minimalistic phenomenological model describes the
pseudoscalar liquid crystal phase of DNA and the phase transition
phenomena. We point out to a possibility that a pseudoscalar liquid
nano-crystal can be assembled via DNA-programming.
\end{abstract}

\pacs{87.14.gk, 87.15.Zg, 61.46.-w, 64.70.Nd, 64.70.pp, 64.60.Cn}

\maketitle

DNA is a biopolymer that provides with the basic mechanisms of
biological information processing. Schr\"{o}dinger was the first
who recognized that a gene or the whole chromosome thread is an
aperiodic crystal \cite{schroedinger}. It consists of two types of
base pairs (AT and CG nucleotide pairs) which are placed in a
random order along the molecular chain. Meanwhile natural DNA
exists as a double helix bound state \cite{watson}. Each base in
one strand is connected to a base in the other strand by hydrogen
bond link. Successive bonds are slightly angled to each other
leading to the famous double-helix form of DNA. Depending on its
environment and mechanical strain, the DNA double helix may have
several different helical forms, of which the most common are
right-handed $A$- and $B$-DNA \cite{alberts}, and the most
peculiar is the left-handed $Z$-DNA \cite{z-dna}. Several
different helical conformations of DNA may even coexist as domains
of the same long molecule \cite{ha} as well as transform into each
other.

Previously it was pointed out \cite{grosberg} that different
condensed matter phases may be formed inside long polymers,
including a liquid crystal phase. In the present paper we will
concentrate on the formation of a novel liquid crystal phase
within a long DNA molecule. The main idea here is that the pieces
of the molecule may have different helicity (being in $B$ or $Z$
forms). This situation is similar to the case of the {\it
pseudoscalar liquid crystal} that was hypothetically introduced by
Zel'dovich back in 1974 \cite{zeldovich}. The peculiarity of such
a phase is that a local right-handed helix may transform into a
left-handed one and vice versa meanwhile interacting with other
helical parts of the molecule. Change of the temperature leads to
an ordering that happens via a spontaneously broken chiral
symmetry phase transition. The ordered phase thus represents the
pseudoscalar liquid crystal \cite{zeldovich}. The order parameter
for liquid crystals is the orientation of its molecules and for a
pseudoscalar liquid crystal the order parameter is the chirality
\cite{pseudoscalar}. In our case it is the local difference
between the number of oppositely-handed helices. A substantial
difference from the other liquid crystals \cite{liquid crystals}
consisting of chiral molecules is that the helices in pseudoscalar
crystals can change their chirality under external conditions.

Consider a DNA molecule as a sequence of pieces containing right-
and left-handed helices, being in one of the two $B$ or $Z$ forms
with opposite chiralities. The pieces interact with each other that
may lead, as we show below, to phase transitions bringing the
molecule, {\it e.g.}, to a pure left-handed $Z$-DNA. The proposed
mechanism can explain how one of the structures, say the $Z$-form,
emerges during intracellular processes. For the quantitative
purposes we will employ a combination of Ising-type nearest-neighbor
interactions and random interactions of distant along the chain
parts which may come close. The long-range interactions may occur
due to random bending of the DNA molecule caused, {\it e.g.}, by
looping proteins \cite{protein-mediated loops}.

We now present the following simple phenomenological model for
description of pseudoscalar liquid crystal phase of DNA. This is a
statistical mechanical model defined on a one-dimensional lattice
with each site corresponding to a rung of the ladder. A spin
variable $\chi_i$ is associated with each site $i$, where we choose
$\chi_i=-1$ to correspond to the energetically more favorable
right-handed helical rotation (that means the particular base pair
belongs to the $B$-form segment of the molecule) and $\chi_i = +1$
for the base pair in the left-handed $Z$-form. Notice that
one-dimensional Ising models with chiral variables were applied to
study random copolymers and bent-core liquid crystals in
\cite{selinger}. Their study can be easily extended to account for
long-range interactions, e.g., via loop formation in the case of
copolymers.

It is reasonable to assume that the neighbor base pairs will try to
get into the same form and thus we will take the nearest-neighbor
interactions to be ferromagnetic. Besides the nearest-neighbor
interactions we will take into account random interactions between
base pairs which are distant along the chain but come close in
space. We assume an arbitrary folding of the DNA molecule so that
any two base pairs may get connected, e.g., via the looping proteins
\cite{protein-mediated loops}. The proposed model has the following
Hamiltonian
\begin{equation}
H= - g\sum_{i=1}^{N-1}\chi_i\chi_{i+1} -\gamma\sum_{i<j}^N J_{ij}
\chi_i \chi_j - h\sum_{i=1}^N\chi_i + \alpha\sum_{i<j}^N J_{ij}
\label{dna}
\end{equation}
where $g>0$ is the coupling parameter of nearest-neighbor
interactions responsible for the torsional stiffness of the polymer;
$J_{ij}$ are the link variables, taking values $0$ and $1$ when the
$i$ and $j$ nodes are uncoupled or coupled, {\it e.g.}, by looping
proteins, correspondingly; $\gamma>0$ is the energy of interaction
between two distant base pairs coupled via appeared link; $h$ is the
half of the energy difference between the states with opposite
helicity; $\alpha$ is the energy of formation of a link connecting
$i$ and $j$ sites. We introduce and will use another parameter $c$
defined via ${c}/{(N-c)} = e^{-\alpha\beta}$, where $\beta$ is the
inverse temperature. The ratio can be roughly treated as the
probability of a link formation (see \cite{statnets} for more
rigorous formulations and details of a related model that describes
a network of fluctuating links). We will assume sparse connectivity
$\frac{c}{N} \ll 1$ as we suppose that the number of uncoupled
distant base pairs that tend to get coupled is much less than the
number of base pairs $N$.

The proposed model can be reduced to an effective mean-field
Hamiltonian with the link variables being eliminated. Indeed, we
proceed with calculation of the partition function $Z={\rm Tr}_\chi
{\rm Tr}_J \exp\left[-\beta H \right]$, where ${\rm
Tr}_\chi\equiv\prod_{i=1}^N \sum_{\chi_i=\pm 1}$ and ${\rm
Tr}_J\equiv\prod_{ij}^N\sum_{ J_{ij}=0,1}$ are traces over the
corresponding variables. Taking the trace ${\rm Tr}_J$ we obtain the
following expression for the partition function
\begin{eqnarray}
Z = {\rm Tr}_\chi e^{\beta \left[g \sum_{i=1}^{N-1}
\chi_i\chi_{i+1}+h\sum_{i=1}^N\chi_i\right]}
\nonumber \\
\times \prod_{ij}^N \left(1+e^{-\beta\alpha
+\beta\gamma\chi_i\chi_j}\right) \nonumber
\end{eqnarray}
Recalling the definition of $c$ and that $c \ll N$, we get for the
last product factor $\exp\left[\frac{c}{N}\sum_{ij}^N e^{\beta\gamma
\chi_i\chi_j} \right]$. Noting then the identity ($\tau=\pm 1$)
$e^{\beta\gamma \tau}=b_0+b_1\tau$,
$b_r=\tanh^r(\beta\gamma)\cosh(\beta\gamma)$ ($r=0,1$) we can now
rewrite the partition function as
\begin{eqnarray}
Z = e^{\frac{1}{2}c(N-1)\cosh\beta\gamma} \cdot {\rm Tr}_\chi
e^{\beta
\left[g\sum_{i=1}^{N-1}\chi_i\chi_{i+1}+h\sum_{i=1}^N\chi_i\right]} \nonumber \\
\times \exp [\frac{c}{N}\sinh\beta\gamma \sum_{ij}^N \chi_i \chi_j]
\nonumber
\end{eqnarray}
Thus we have arrived at the partition function with the following
effective Hamiltonian
\begin{eqnarray}
H = - g\sum_{i=1}^{N-1}\chi_i\chi_{i+1} - \gamma'\sum_{i<j}^N \chi_i
\chi_j - h\sum_{i=1}^N\chi_i \label{dna}
\end{eqnarray}
where the coupling parameter is
$\gamma'=\frac{c}{N}\sinh\beta\gamma$. The first term is
responsible for interactions between nearest-neighbor base pairs.
It assures that neighbor pairs tend to get the same value of
chirality. The second term in the Hamiltonian describes the
effective mean-field interaction between base pairs. The last term
is the half of the energy difference between the right- and
left-handed helices at the sites of the chain.

\begin{figure}
\includegraphics[width=1.05\linewidth,angle=0]{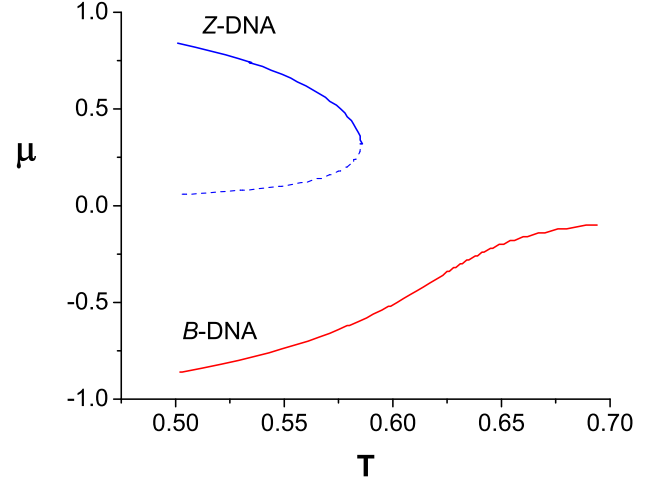}
\hfill \caption{Chirality vs temperature. The parameters are $g =
0.2$ kcal/mol, $\gamma = 0.02$ kcal/mol, $c = 10$ and $h = 0.01$
kcal/mol. There is a second-order phase transition at the critical
temperature $T_c\approx 0.6$ kcal/mol.} \label{fig1}
\end{figure}

We now introduce the order parameter
$\mu=\frac{1}{N}\sum_{i=1}^N\chi_i$ which takes the values $\mu =
-1$ and $\mu = 1$ for the pure samples of $B$ and $Z$ forms of DNA,
respectively. To calculate the partition function for the above
mean-field Hamiltonian we use the relationship $\sum_{i<j}^N \chi_i
\chi_j = \frac{1}{2}(\sum_{i=1}^N \chi_i)^2 - \frac{1}{2}N$, the
Hubbard-Stratonovich transformation $e^{\frac{1}{2}a(\sum_{i=1}^N
\chi_i)^2}=\int^{+\infty} _{-\infty} \frac{d \mu}{\sqrt{2\pi/a}}
e^{-\frac{1}{2}a\mu^2 + a\mu\sum_{i=1}^N \chi_i}$ and the expression
for the partition function of the one-dimensional (1D) Ising model
\cite{lavis}. Then the partition function takes the form $Z \propto
\int^{+\infty} _{-\infty} d\mu e^{- \beta N f(\mu)}$ with the
effective free energy $f(\mu)$ given by
\begin{eqnarray}
f(\mu) = \frac{1}{2}b \mu^2 - \beta^{-1}\ln [\cosh \beta (h + b \mu) + \nonumber \\
\sqrt{\sinh^2 \beta (h + b \mu) + e^{-4\beta g}}] \label{fenergy}
\end{eqnarray}
where we have defined $b = c\sinh\beta\gamma$.

To further analyze equilibrium properties of the substance under
consideration the extreme values of $\mu$ are to be obtained via
$f'(\mu)=0$ that leads to the following state equation
\begin{eqnarray}
\mu = \frac{\sinh\beta (h + b \mu)}{\sqrt{\sinh^{2}\beta (h + b
\mu) + e^{-4\beta g}}} \label{mean}
\end{eqnarray}
The dependence of the order parameter $\mu$ on temperature $T$ is
presented in Fig.\ref{fig1}. For the zero-field case ($h=0$) one
would have a second-order phase transition. The molecule
spontaneously transforms into a structure that has non-zero total
chirality. It chooses between $B$ and $Z$ forms. The critical
temperature of the phase transition is determined via $\beta_c
b\cdot e^{2\beta_c g} = 1$. If the field is non-zero then the system
goes through the phase transition to get the form that is favored by
the sign of the field. The field $h$ depends on the external
parameters that are determined by environmental conditions such as
pH value, salt concentration, and other chemical as well as
mechanical factors (e.g., locally applied forces and torques).

\begin{figure}
\includegraphics[width=1.05\linewidth,angle=0]{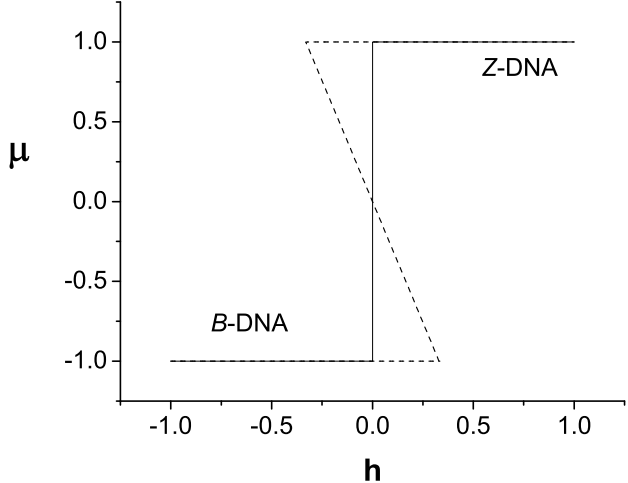}
\hfill \caption{Chirality vs the base pair binding energy for the
parameters $g = 8.5$ kcal/mol, $\gamma = 0.02$ kcal/mol, $c = 10$,
and $k_B T = 0.6$ kcal/mol. The first-order $B-Z$ phase transition
occurs at the critical value $h_c=0$. Dashes indicate the metastable
states.} \label{fig2}
\end{figure}

For a fixed constant temperature there can be a first-order phase
transition for chirality versus the field as demonstrated in
Fig.\ref{fig2}. The phase transition transforms the right-handed
$B$-DNA into left-handed $Z$-DNA at the critical point $h_c=0$
where the external parameter changes its sign. $Z$-DNA corresponds
to a state of the molecule with the energy higher than the one in
$B$ form. Among the factors which induce $B-Z$ transitions are
$(K-X)_n-K$ peptide, polynuclear $Pt$-complexes, etc
\cite{fuertes}. The inverse $Z-B$ transition can be induced by
KWGK peptide, daunomycin, and several other substances
\cite{fuertes}.

$B-Z$ transitions have been extensively investigated that led to
several models \cite{fuertes}. Recently molecular dynamics
\cite{kastenholz} and coarse-grained model \cite{lim} simulations
were carried out to provide with more satisfactory explanation of
the molecular mechanisms of the $B-Z$ transitions. In contrast, we
here focus on the phenomenology of the transitions that are
described by the minimalistic model presented above. The main goal
was to demonstrate the possibility of transformations under the
influence of the external factors and temperature.

As we mentioned it was shown that different helical conformations
of DNA may coexist as domains of the same long molecule \cite{ha}.
Now consider a chromatin, that is the complex of DNA and proteins in
which the generic material is packaged inside the cells with nuclei \cite{alberts}.
Let us assume that a DNA molecule in chromatin happens to consist
of right- and left-handed parts placed in a random order along the
chain. Besides, chromatin structure is dynamic \cite{felsenfeld} that
is important because changes in its structure can be inherited independent
of the DNA sequence itself, the so-called epigenetic inheritance. Our model
is able, in principle, to mimick the case of DNA molecule with a dynamic structure
as it provides with a plasticity mechanism via the fluctuating links
(random bending of DNA molecule caused by looping proteins) that (dis)connect base pairs.
Thus a question arises if a DNA molecule in a chromatin can naturally exist
in the state of matter we consider here, the pseudoscalar liquid crystal.
This is truly an astonishing possibility and further experimental and theoretical
investigation of the molecular details would shed more light onto that.

In conclusion, we have shown that a long DNA molecule, that consists of
aperiodically ordered right-handed $B$ and left-handed $Z$ fragments, can form a novel condensed phase
of matter, the pseudoscalar liquid crystal. We presented a minimalistic phenomenological
model that describes the pseudoscalar liquid crystal phase of DNA and related phase transition
phenomena. Generally speaking, one can use the following 1D Hamiltonian for theoretical
treatment of the pseudoscalar liquid crystal
\begin{eqnarray}
H = \frac{g}{2} \int dl \left(\frac{\partial \chi(l)}{\partial l}\right)^2 + \gamma \int\int dl dl' J(l,l') \chi(l) \chi(l') \nonumber \\
+ \int dl h(l) \chi(l) \nonumber
\end{eqnarray}
with the continuous along the chain variable $\chi(l)$ and the interaction potential $J(l,l')$. Here $\chi(l)$ plays the role
of the liquid crystal chiral variable defined as $\textbf{n} \cdot \nabla \times \textbf{n}$ with $\textbf{n}$ being the director for conventional liquid crystals
\cite{liquid crystals}, and $h(l)$ being a heterogenous (random) external field.

Further experimental research is needed to identify if this condensed state of matter
exists in nature. Yet another foreseeable way towards the new state of matter would be assembling
pseudoscalar liquid nano-crystals via DNA-programming \cite{mirkin,sleiman}.

We thank A. E. Allahverdyan, D. Mukamel, R.A. Roemer, E. I. Shakhnovich, and M.
C. Williams for comments and discussions. This work was supported by National Science
Council in Taiwan under Grant Nos. NSC 96-2911-M 001-003-MY3, NSC 96-2811-M-001-018
and NSC 97-2811-M-001-055, by National Center for Theoretical Sciences in Taiwan, and by
Academia Sinica (Taiwan) under Grant No. AS-95-TP-A07.


\begin{thebibliography}{99}
\bibitem{schroedinger}E. Schr\"{o}dinger, {\it What is life: The physical aspect of living
cell} (University Press, Cambridge, 1944).

\bibitem{watson}J.D. Watson and F.H.C. Crick, Nature {\bf 171}, 737
(1953).

\bibitem{alberts}B. Alberts et al, {\it Molecular Biology of the Cell}
(4th edn, Garland Science, New York, 2002).

\bibitem{z-dna}A. Wang et al, Nature {\bf 282}, 680 (1979);
A. Rich and S. Zhang, Nature Rev. Genet. {\bf 4}, 566 (2003); P.C.
Champ, S. Maurice, J.M. Vargason, T. Camp, and P.S. Ho, Nucleic
Acids Res. {\bf 32}, 6501 (2004).

\bibitem{ha}S.C. Ha, K. Lowenhaupt, A. Rich, Y.G. Kim, and K.K. Kim,
Nature (London) {\bf 437}, 1183 (2005).

\bibitem{grosberg}A.Yu. Grosberg and A.R. Khokhlov, {\it Statistical Physics of
Macromolecules}, Chap. 3, {\bf \S}22 (AIP Press, Woodbury NY,
1994).

\bibitem{zeldovich}Ya.B. Zel'dovich, Zh. Eksp. Teor. Fiz. {\bf 67}, 2357 (1974)
[Sov. Phys. JETP {\bf 40}, 1170 (1975)].

\bibitem{pseudoscalar}A pseudoscalar is a quantity that behaves like a scalar,
except that it changes sign under a parity inversion. The examples
of pseudoscalars are magnetic flux, spin helicity and the
chirality which is considered in the present paper.

\bibitem{liquid crystals}S. Chandrasekhar, {\it Liquid Crystals} (Cambridge
University Press, New York, 1977); P.G. de Gennes and J. Prost,
{\it The Physics of Liquid Crystals} (Clarendon Press, Oxford,
1993).


\bibitem{protein-mediated loops}L. Saiz and J.M.G. Vilar, Curr. Opin. Struct. Biol. {\bf 16}, 344 (2006);
G.A. Maston, S.K. Evans, and M.R. Green, Annu. Rev. Genomics Hum.
Genet. {\bf 7}, 29 (2006); H.G. Garcia, P. Grayson, L. Han, M.
Inamdar, J. Kondev, P.C. Nelson, R. Phillips, J. Widom, and P.A.
Wiggins, Biopolymers {\bf 85}, 115 (2007).

\bibitem{selinger}J.V. Selinger and R.L.B. Selinger, Phys. Rev.
Lett. {\bf 76}, 58 (1996); J.V. Selinger, Phys. Rev. Lett. {\bf
90}, 165501 (2003).

\bibitem{statnets}A.E. Allahverdyan and K.G. Petrosyan, Europhys. Lett. {\bf 75}, 908 (2006).

\bibitem{lavis}D.A. Lavis and G.M. Bell, {\it Statistical mechanics of lattice
systems}, v. 1 (Springer-Verlag, Berlin, 1999).

\bibitem{fuertes}M.A. Fuertes, V. Cepeda, C. Alonso, and J.M.
Perez, Chem. Rev. {\bf 106}, 2045 (2006).

\bibitem{kastenholz}M.A. Kastenholz, T.U. Schwartz, and P.H.
H\"{u}nenberger, Biophys. J. {\bf 91}, 2976 (2006).

\bibitem{lim}W. Lim, Phys. Rev. E {\bf 75}, 031918 (2007).

\bibitem{felsenfeld}G. Felsenfeld and M. Groudine, Nature {\bf 421}, 448 (2003).

\bibitem{mirkin}D. Nykypanchuk, M.M. Maye, D. van der Lelie, and O. Gang, Nature {\bf
451}, 549 (2008); S.Y. Park, A.K.R. Lytton-Jean, B. Lee, S.
Weigand, G.C. Schatz, and C.A. Mirkin, Nature {\bf 451}, 553
(2008).

\bibitem{sleiman}F.A. Aldaye, A.L. Palmer, and H.F. Sleiman, Science
{\bf 321}, 1795 (2008).

\end{thebibliography}
\end{document}